\documentclass[]{spie}  

\usepackage{amsmath,amsfonts,amssymb}
\usepackage{graphicx}
\usepackage[colorlinks=true, allcolors=blue]{hyperref}
\usepackage{cleveref}
\usepackage{subfigure}
\usepackage{float}

\title{Phase-Retrieval-Based Wavefront Metrology for High Contrast Coronagraphy}

\author{Gregory R. Brady\supit{a}, Christopher Moriarty\supit{a}, Peter Petrone\supit{a}, Iva Laginja\supit{a}, Keira Brooks\supit{a}, Tom Comeau\supit{a}, Lucie Leboulleux\supit{a}\supit{b}\supit{c}, R\'{e}mi Soummer\supit{a}
\skiplinehalf
\supit{a} Space Telescope Science Institute, 3700 San Martin Drive, Baltimore, MD 21218, USA\\
\supit{b} Aix Marseille Universit\'{e}, CNRS, LAM (Laboratoire d\'{}Astrophysique de Marseille) UMR 7326, 13388, Marseille, France\\
\supit{c} Office National d\'{}Etudes et de Recherches A\'{e}rospatiales, 29 Avenue de la Division Leclerc, 92320 Ch\^{a}tillon, France\\
}

\authorinfo{Further author information: send correspondence to Gregory Brady: E-mail: gbrady@stsci.edu, Telephone: +1 410 338 6504}

\begin{document} 
\maketitle

\begin{abstract}
We discuss the use of parametric phase-diverse phase retrieval as an \textit{in-situ} high-fidelity wavefront measurement method to characterize and optimize the transmitted wavefront of a high-contrast coronagraphic instrument. We apply our method to correct the transmitted wavefront of the HiCAT (High contrast imager for Complex Aperture Telescopes) coronagraphic testbed.  This correction requires a series of calibration steps, which we describe.  The correction improves the system wavefront from 16 nm RMS to 3.0 nm RMS.
\end{abstract}


\section{Introduction}
\label{sec:Introduction}
The next generation of space telescopes for direct imaging and spectroscopy of exoplanets will require coronagraphs with high-contrasts, on the order of $10^{-10}$.  To achieve this, iterative speckle nulling procedures will be used with science data to eliminate residual light passing through the coronagraph by perturbing a deformable mirror (DM) and correcting the end-to-end system wavefront on the scale of a few picometers.  However, most of these procedures assume a linear approximation when relating the measured intensity to the wavefront phase, which becomes invalid for phase distributions with an amplitude larger than a small fraction of the wavelength.\cite{Amir:2007,Bottom:2016,2017JATIS...3d9002M}
Thus we desire a method for decreasing the coronagraph system's wavefront error so that it is within the  valid range of the linear approximation.

In this work we describe a procedure for measuring the wavefront of a well-aligned system using a phase retrieval algorithm based on an explicit optimization of parameters describing the phase and other pertinent system characteristics.  Phase retrieval algorithms accurately model the propagation from the pupil plane to image plane and do not make use of a linear approximation.   The wavefront measured using phase retrieval is used to derive a correction that is applied to the DM.  Phase retrieval is not sufficiently sensitive to perform measurements on the picometer scale and is used to pre-correct the system wavefront error to the order of a few nanometers prior to speckle nulling. 

In addition, we will describe a series of phase-retrieval-based calibration steps that are necessary to ensure that the wavefront correction is properly registered to the DM.  Finally, we describe our final measurement of the system wavefront demonstrating dramatic improvement, bringing the wavefront error well within the valid range of the linear phase approximation.
\section{Parametric Phase Retrieval Algorithm}
\label{sec:Parametric Phase Retrieval Algorithm}
Phase retrieval algorithms typically infer the phase in the exit pupil of an optical system using one or more measurements of the intensity patterns that the system produces in a plane near its nominal focus.\cite{Fienup:78}  A parametric phase retrieval algorithm specifies a system to be considered using a set of parameters and uses an optimization algorithm to find values of these parameters that gives a minimum value for an objective function (also termed an error metric). The parameters used typically specify the phase in the exit pupil using a polynomial expansion, such as Zernike polynomials, or another basis set, such as family of delta-functions to specify phase at individual pixel locations.\cite{Fienup:82,Fienup:99}  Other parameters can make the algorithm more robust by allowing it to correct for inexactly known system parameters, such as the distance from the exit pupil to the measurement plane or planes,\cite{bradyPhD2008, Thurman:09_amplMetrics}the centration of the measured spots on the detector array,\cite{Brady:09,Thurman:09_amplMetrics} the amplitude distribution in the pupil plane,\cite{Brady:06_CmplxPupil,Thurman:09_cmplxPupil} or the sampling rate of the image data.\cite{Jurling:14}  Further, these algorithms have the ability to jointly optimize over multiple images, allowing for a more robust estimate of the wavefront.  This ability to compensate for systematic uncertainties make a parametric, optimization-based algorithm an attractive choice when compared to iterative-transform-type algorithms such as the Gerchberg-Saxton-Misell algorithm.\cite{Gerchberg72,Misell_1973,Fienup:82,dean2006}  Iterative transform algorithms are less flexible and less robust (i.e. more prone to stagnating in local minima) and rely on properties of the algorithm alone for their convergence characteristics.  

\subsection{Objective Function}
Here we employ a bias and gain-insensitive objective function to characterize the agreement between our modeled and measured intensity patterns\cite{Thurman:09_amplMetrics, Thurman:09_biasInsens}, 
\begin{equation}
    \label{eq:ObjFun}
    \Phi = 1 - \frac{1}{K} \sum_{k}^{K} \frac{\left[\sum\limits_{p,q}W_k(p,q)G_k(p,q)\widehat{G}_k(p,q)\right]^2}{\left[\sum\limits_{p,q}W_k(p,q)G^2_k(p,q)\right]\left[\sum\limits_{p,q}W_k(p,q)\widehat{G}^2_k(p,q)\right]},
\end{equation}
where $p$ and $q$ are sample indices, $G_k(p,q)$ is one of $K$ measured intensity patterns, $\widehat{G}_k(p,q)$ is the corresponding calculated intensity pattern and $W_k(p,q)$ is a weighting function used to ignore the contribution of bad pixels or to favorably weight certain portions of the intensity pattern with, for example, better Signal-to-Noise ratio (SNR). Throughout this paper we use the convention that variables topped with a circumflex, such as $\widehat{G}_k(p,q)$, are being estimated while other values are known through measurement or convention. If $G_k(p,q)$ and $\widehat{G}_k(p,q)$ are identical, the value of the objective function is zero.

\subsection{Phase Model}
We typically model the field, $\widehat{E}_p(m,n)$, in the exit pupil using one of two methods.  The first method is by simply using sampled representations of the amplitude $\widehat{A}\left(m,n\right)$ and phase $\widehat{\phi}\left(m,n\right)$ so that we have
\begin{equation}
    \widehat{E}_p\left(m,n\right) = \widehat{A}\left(m,n\right)\exp\left[i\widehat{\phi}\left(m,n\right)\right]
\end{equation}
at sample indices $\left(m, n\right)$.  We term this representation as point-by-point.

The second method is by representing the phase by the coefficients $\widehat{\alpha}_j$ of a basis set such as the Zernike polynomials $Z_j\left(m,n\right)$ so that we have
\begin{equation}
    \widehat{E}_p\left(m,n\right) = \widehat{A}\left(m,n\right)\exp \left[ i\sum\limits_{j}\widehat{\alpha}_jZ_j\left(m,n\right) \right].
\end{equation}
It is common practise to initially optimize over a few low order Zernike terms before continuing the optimization with higher order Zernikes or point-by-point phase values.

\subsection{Propagation Model}
We employ a two step propagation to calculate the estimated field in each of the measurement planes from the field in the pupil plane.  First, Fresnel diffraction is used to calculate the field $\widehat{E}_f(p,q)$ in the paraxial focal plane,
\begin{equation}
    \widehat{E}_f(p,q) = \frac{1}{N}\exp\left[\frac{i\pi}{\lambda f}\left(p^2\Delta^2_p +q^2\Delta^2_q\right)\right]\sum\limits_{m,n}^N\widehat{E}_p(m,n)\exp\left[\frac{-i2\pi}{N}\left(mp + nq\right)\right],
\end{equation}
where $\lambda$ is the wavelength, $f$ is the distance between the exit pupil and paraxial focal plane, $N$ is the size of the square arrays representing the pupil and focal plane fields, and $\left(\Delta_p,\Delta_q\right)$ are the spacing between samples (in units of length) in the paraxial focal plane.  This is typically equivalent to the pixel spacing of the detector used to record the image data.
The second step uses angular spectrum propagation to propagate the small distance between the paraxial focal plane and each of the slightly defocused measurement planes.  We calculate the angular spectrum of plane waves $\widehat{U}_f\left(m,n\right)$ in the paraxial focal plane using a Fourier transform
\begin{equation}
    \widehat{U}_f\left(m,n\right) = \frac{1}{N}\sum\limits_{p,q}^N \widehat{E}_f(p,q) \exp\left[-i\frac{2\pi}{N}\left(pm + nq\right)\right].
\end{equation}
The angular spectrum is then propagated to each of the $K$ measurement planes,
\begin{equation}
    \widehat{U}_k\left(m,n\right) = \widehat{U}_f\left(m,n\right)  \exp\left(i2\pi\widehat{z}_k\sqrt{\frac{1}{\lambda^2}-m^2\Delta^2_m - n^2\Delta^2_n}\right),
\end{equation}
where $\left(\Delta_m,\Delta_n\right)$ are frequency domain sample spacings (in units of inverse length) given by
\begin{equation}
    \Delta_{m,n} = \frac{1}{N\Delta_{p,q}}
\end{equation}
and $\widehat{z}_k$ is the distance to the \textit{k}\textsuperscript{th} measurement plane. Note that we have included a circumflex in $\widehat{z}_k$ as we typically include this as a free parameter in the optimization to compensate for possible positioning errors.  The propagated field in the spatial domain is computed from the angular spectrum using an inverse Fourier transform,
\begin{equation}
    \widehat{E}_k\left(p,q\right) = \frac{1}{N}\sum\limits_{m,n}^N \widehat{U}_k(m,n) \exp\left[i\frac{2\pi}{N}\left(mp + nq\right)\right].
\end{equation}
The intensity in the measurement plane is then determined from the field,
\begin{equation}
    \widehat{I}_k\left(p,q\right) = \left| \widehat{E}_k\left(p,q\right) \right|^2.
\end{equation}
A final step is often performed on this intensity image to shift the intensity spot laterally to model the effect of a detector array not centered at the origin.  This is done on a sub-pixel scale using the Fourier shift theorem.  First, we perform the forward transform
\begin{equation}
    \widehat{f}_k\left(m,n\right) = \frac{1}{N}\sum\limits_{p,q}^N \widehat{I}_k(p,q) \exp\left[-i\frac{2\pi}{N}\left(mp + nq\right)\right],
\end{equation}
then we multiply by the linear phase Fourier filter
\begin{equation}
    \widehat{g}_k\left(m,n\right) = \exp\left[\frac{-i2\pi}{N}\left(m\widehat{x}_k + n \widehat{y}_k\right)\right]\widehat{f}_k\left(m,n\right),
\end{equation}
where $\left(\widehat{x}_k,\widehat{y}_k\right)$ is the amount by which the image is shifted in pixel units.  Note that these are also typically optimized for.  Inverse transforming this gives us our final modeled intensity distribution,
\begin{equation}
    \widehat{G}_k\left(p,q\right) = \frac{1}{N}\sum\limits_{m,n}^N \widehat{g}_k\left(m,n\right) \exp\left[i\frac{2\pi}{N}\left(mp + nq\right)\right],
\end{equation}
which, along with the measured intensity data $G_k\left(p,q\right)$, is used to calculate the error metric given in \autoref{eq:ObjFun}.
\subsection{Optimization}
We typically employ gradient-based algorithms to minimize the objective function, such as the conjugate gradient\cite{Press:2007:NRE:1403886} or limited-memory BFGS\cite{10.2307/2003239,doi:10.1137/0916069,Zhu:1997:ALF:279232.279236} algorithms.  These algorithms are most useful when efficient analytic expressions for the derivatives of the objective function with respect to the system parameters are available, rather than computing the derivatives using finite differences.  This is particularly true in the case of point-by-point estimates of the phase, where $N \times N$ derivatives must be calculated and $N$ is typically a few hundred to a few thousand.  Efficient expressions for all of the parameters discussed have been derived and are compiled in the appendix of Reference \cite[]{Thurman:09_amplMetrics}.  To reiterate, our phase retrieval code has the capability to optimize over
\begin{itemize}
    \item Lateral spot position in each phase-diverse image, $\widehat{x}_k$, $\widehat{y}_k$
    \item Axial location of each phase-diverse image, $\widehat{z}_k$
    \item Pupil amplitude distribution, $\widehat{A}\left(m,n\right)$
    \item Pupil phase expressed as point-by-point phase values, $\widehat{\phi}\left(m,n\right)$
    \item Pupil phase expressed as coefficients of Zernike polynomials (or other basis), $\widehat{\alpha}_j$.
\end{itemize}
These can be used jointly or individually, with the output of a preceding optimization serving as the input for subsequent optimizations, as is necessary for the problem at hand.  In the code used in this work, the following sequence is typically used:
\begin{enumerate}
    \item Preprocess image data to calculate data weight masks, $W_k(p,q)$, and initial guesses for lateral positions from image centroids.
    \item Optimize over lateral positions alone, using guess from centroids.
    \item Optimize jointly over lateral position (using result from step 2), axial position and Zernike coefficients.
    \item Fix values for lateral and axial positions at values determined in step 3, optimize over point-by-point phase values using Zernike-derived phase as initial guess.
    \item Optimize over lateral and axial position and point-by-point phase values using results from step 4.
    \item Optimize over lateral and axial position, point-by-point phase using results from step 5 as initial values, and pupil amplitude using assumed pupil as inital guess.
\end{enumerate}
This multiple-step process allows us to determine the unknown parameters of the system robustly.  If we were to proceed directly to step 6 it is likely that the optimization would stagnate in a local minima.  Steps 4 to 6 are necessary only when point-by-point phase values are desired and may be omitted when Zernikes are all that are needed.
\subsubsection{Data Weights}
The data weights $W_k(p,q)$ are typically derived from the measured data in an attempt to weight the data suitably for the optimization to be done.  Typically they are formed by thresholding the input images so that only the central core of the image is retained.  However, there are regions near the image center where the intensity is below the threshold value.  Furthermore this data is important.  To retain these regions a dilation operation is performed to fill in the mask. The resulting weight mask is highly suitable for retrieving the phase over low spatial frequencies, particularly for determining Zernike coefficients.  This type of mask is often useful to reject noise from low intensity regions of the images, which is often where the SNR is unacceptably low. This type of mask is used in steps 2 and 3 above.

In our application we need to work with data that has significant spatial frequency content, where a data mask retaining only the central core would not recover the phase information of interest.  Further, this high spatial frequency information typically has much less energy than the central core.  If a uniform weighting is used across the frame the central core will still dominate and the high-frequency information will effectively be ignored.  To recover this information we have found that it is necessary to weight the region away from the central core much more heavily than the central core.  For example, we have achieved good results with a weight of 1 in the central core and a weight of 500 outside it.  We employed this type of mask in steps 4 to 6 above.  This weighting requires data that has good SNR in the high-frequency region, which we typically obtain by registering and summing hundreds of individual frames.  

\section{Experiment}
\label{sec:EXPERIMENT}
\begin{figure}[h]
\includegraphics[width=1.0\textwidth]{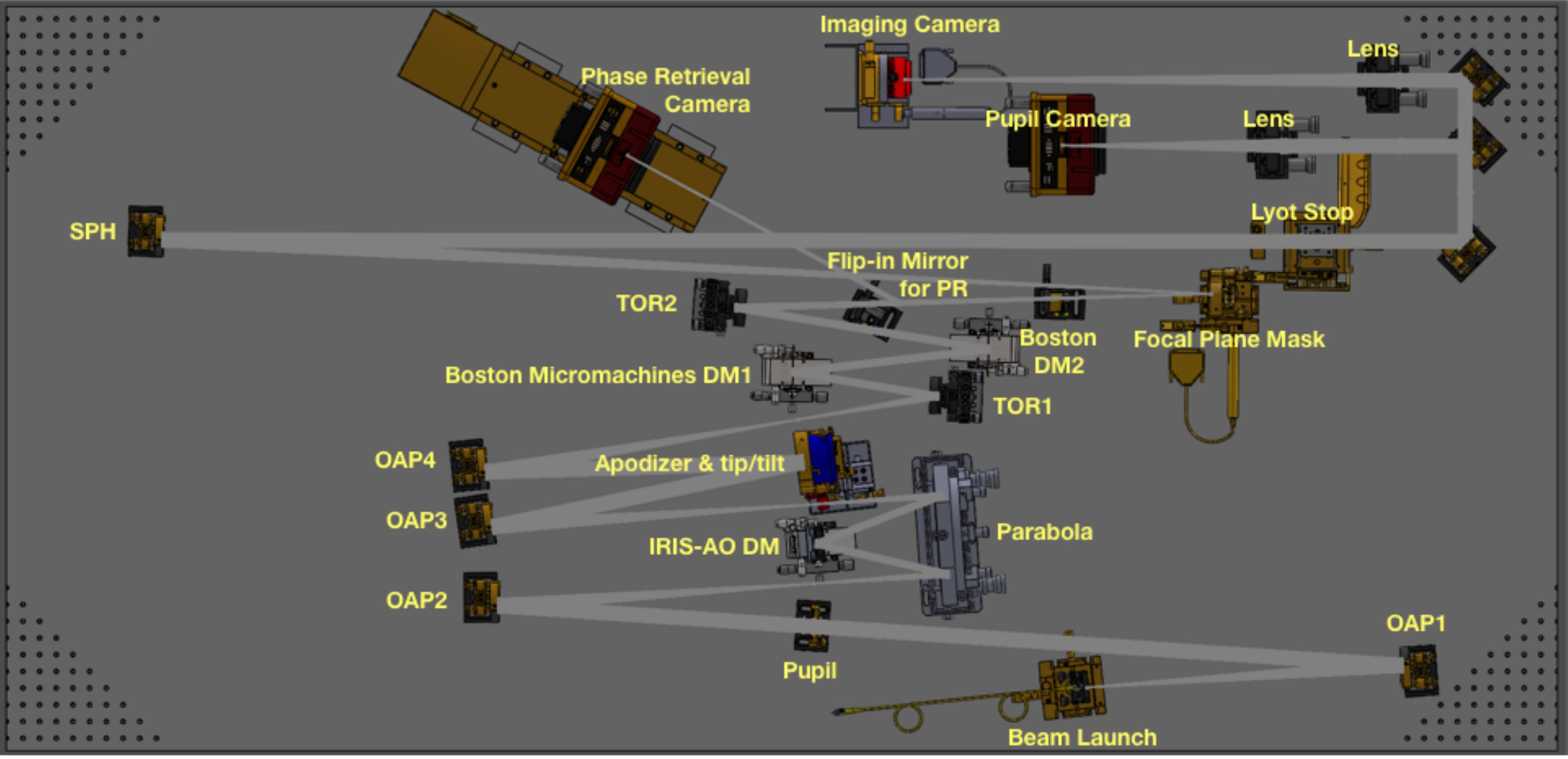}
\caption{HiCAT testbed layout. The testbed includes a telescope simulator (IRIS-AO 37 segment deformable mirror combined with a pupil mask to produce the central obstruction and support structures); an Apodized Pupil Lyot Coronagraph (APLC) combining an apodizer, a hard-edge focal plane mask and a Lyot Stop; Imaging camera including a coronagraphic focal plane, a pupil viewing camera; and a phase retrieval arm to measure the wavefront at the focal plane mask using a high-quality removable fold mirror to minimize non-common path errors. 
\label{fig:hicat_cad} }
\end{figure}

Our goal in this work is to measure, and subsequently correct, the phase of the HiCAT optical system up to the occulting aperture (or Focal Plane Mask, FPM) of the coronagraph.  A sketch of the system is shown in \autoref{fig:hicat_cad}. The system uses a single-mode optical fiber collimated by an Off-Axis Parabolic (OAP) mirror, labeled as OAP1 in \autoref{fig:hicat_cad}, to illuminate the pupil of a telescope simulator.  The telescope simulator consists of a pupil amplitude aperture that is imaged onto an Iris AO hexagonal-segment deformable mirror (IRIS-AO DM) by OAP2 and a parabolic mirror.  This DM allows us to simulate the segments of a segmented aperture telescope.  The DM pupil plane is again reimaged by the parabola and OAP3 onto a reflective apodizer needed for an Apodized Pupil Lyot Coronagraph (APLC).  The apodizer is the first step in controlling the light in the dark zone of the coronagraph.  The apodizer is also used as a tip/tilt mirror to stabilize the beam under servo control.  The apodizer pupil plane is imaged by OAP4 and a toroidal mirror, TOR1, onto a Boston Micromachines continuous face-sheet deformable mirror, DM1.  DM1 is used to correct for residual static aberrations of the system, the aberrations that we are measuring in this work.  The beam reflected from DM1 is directed onto a second identical deformable mirror, DM2.  DM2 is a short distance away and is not exactly conjugate to the other pupils of the system.  Together DM1 and DM2 are used to suppress light in the dark zone of the coronagraph.  After DM2, a second toroidal mirror, TOR2, is used to focus the light to a focus at which the FPM is placed when the coronagraph is operating.  The FPM is responsible for blocking the majority of the star's light.  However, between TOR2 and the FPM, a high-quality mirror can be inserted to intercept the converging beam and direct it to a camera on a motorized stage.  This camera is used to collect phase-diverse phase retrieval data sets, the results of which are our topic here.  A very high-quality point-source image is formed at the FPM, allowing it to block star light most effectively.  Beyond the FPM, a spherical mirror is used to re-image the pupil to the Lyot stop of the coronagraph.  The Lyot stop blocks light diffracted by the hard edge of the FPM that falls outside of the system pupil.  Finally, fold mirrors and lenses are used to form a coronagraphic image with our desired dark zone or a pupil image on the respective cameras.

We have made a small number of notable changes from the nominal arrangement described above to help ensure good  wavefront correction and facilitate phase retrieval measurements using our current software.  First, the nominal pupil aperture has been replaced with a 22 mm diameter aperture to cover the entire area of the Boston Micromachine DMs. This allows us to better correct aberrations that are near the edge of the apodizer pupil (circumscribed diameter of $19.725$ mm, slightly undersized compared to the DM diameter for alignment tolerancing purposes). Because of the apodized profile, we use an 18mm aperture mask as a full-aperture proxy to evaluate the wavefront over the apodizer.  Secondly, the IRIS-AO DM and apodizer have been replaced with surrogate flat mirrors. This allows us to use our current phase retrieval code which does not currently handle the case of a segmented, apodized pupil.  Code is in development to handle this situation.

\subsection{Pupil Calibration}
\begin{figure}[h]
\centering
\subfigure[]{
\label{fig:letterFDM}
\includegraphics[height=2.25in]{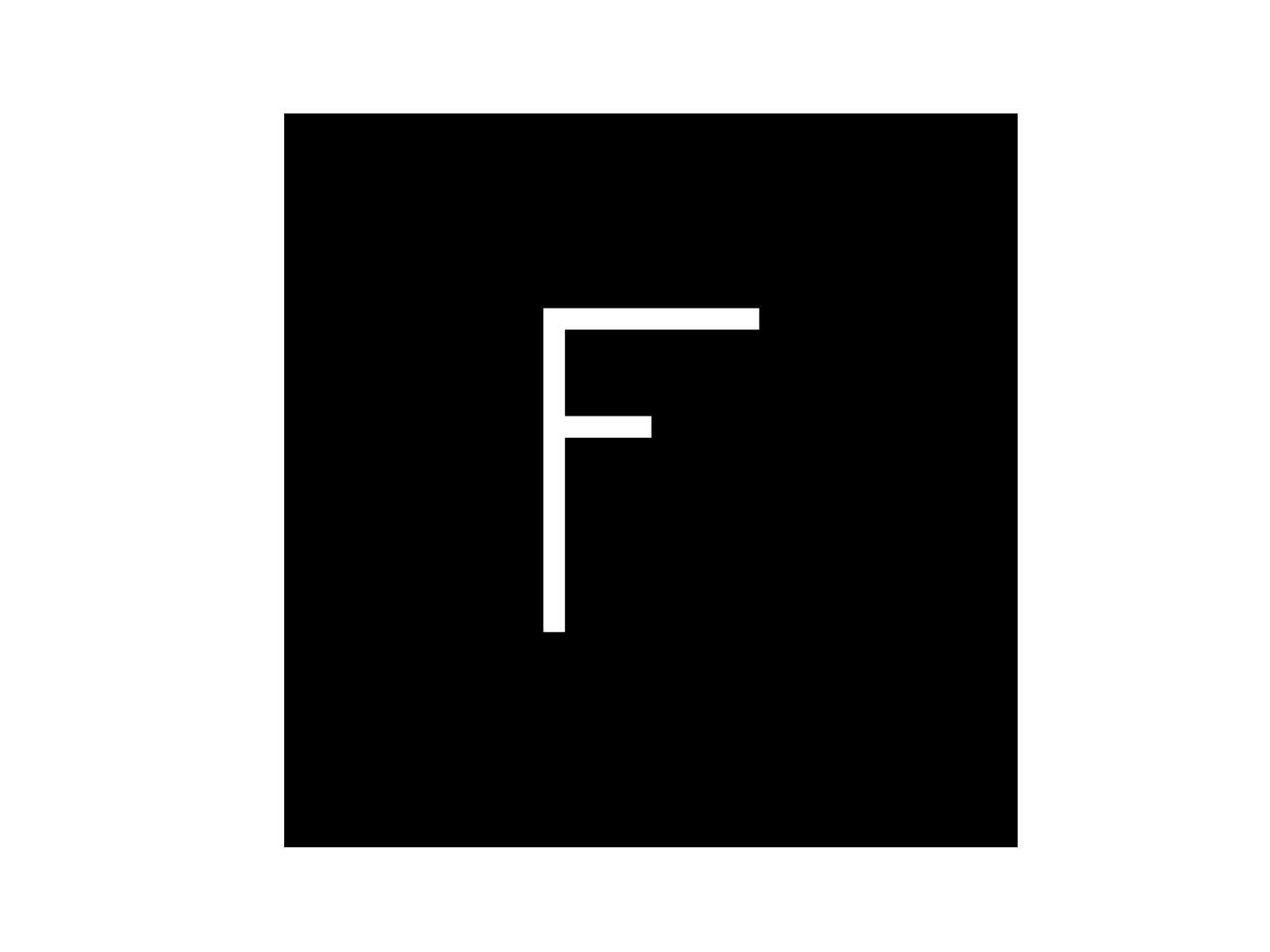}}
\qquad
\subfigure[]{
\label{fig:letterFPhase}
\includegraphics[height=2.25in]{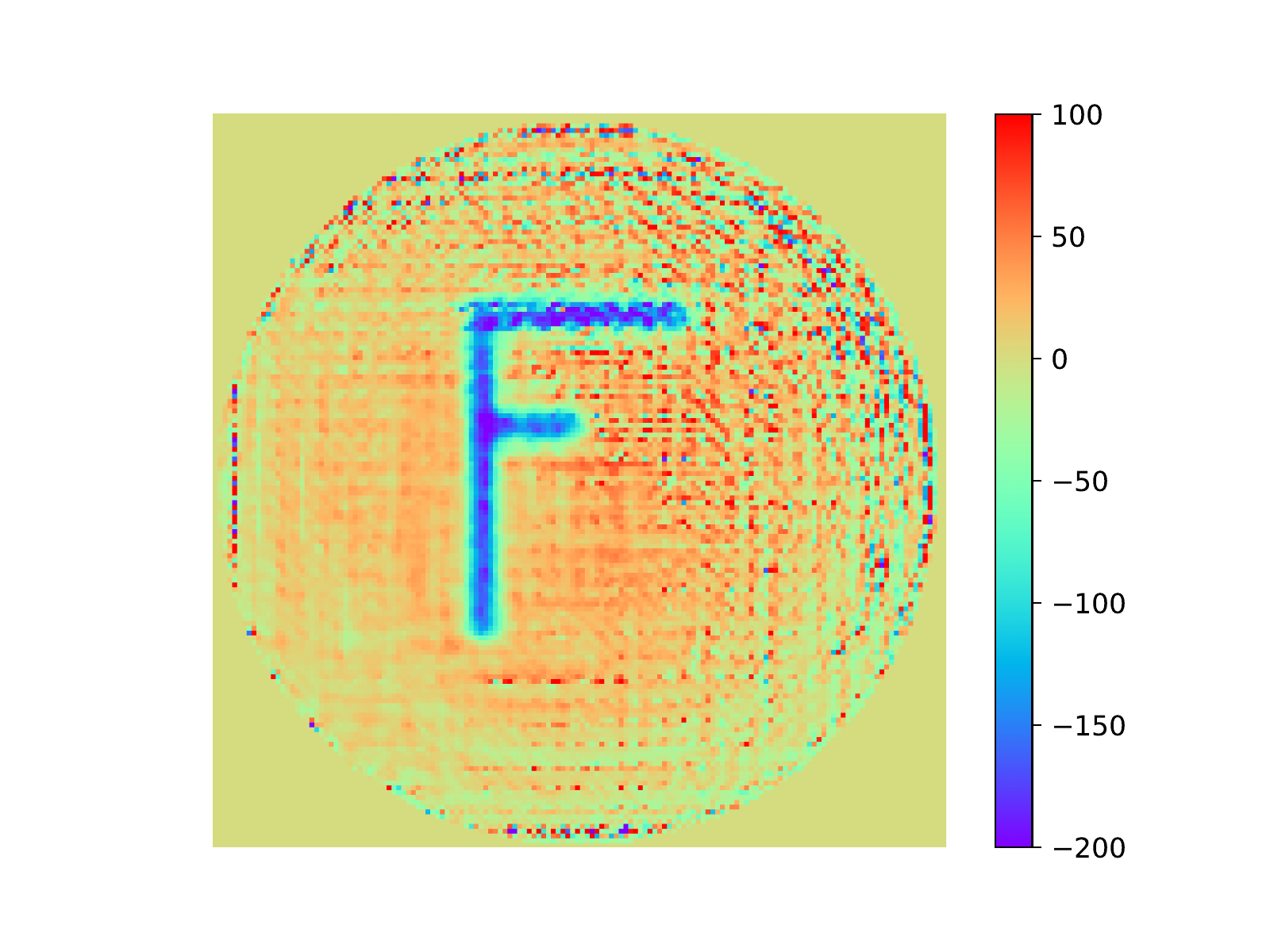}}
\caption{Letter F pattern (a) DM command and (b) retrieved phase.  Color scale is in nm.
\label{fig:letterF} }
\end{figure}

In order to apply the measured phase distributions it is necessary to perform a geometric calibration so that the wavefront can be applied to the DM.  This was achieved by applying calibration phase patterns to the DM and recovering them using phase retrieval.
The first step of the calibration is to apply a letter ``F" pattern to the DM, which allows us to determine how the DM and phase retrieval results are flipped, left-right and up-down, and rotated with respect to each other.  In this step we also confirmed the polarity of the retrieved phase as compared to the voltages applied to the DM.  The DM command and retrieved phase are shown in \autoref{fig:letterF}, with the determined flips and rotations applied.

\begin{figure}[h]
\centering
\subfigure[]{
\label{fig:centerPokeDM}
\includegraphics[height=2.25in]{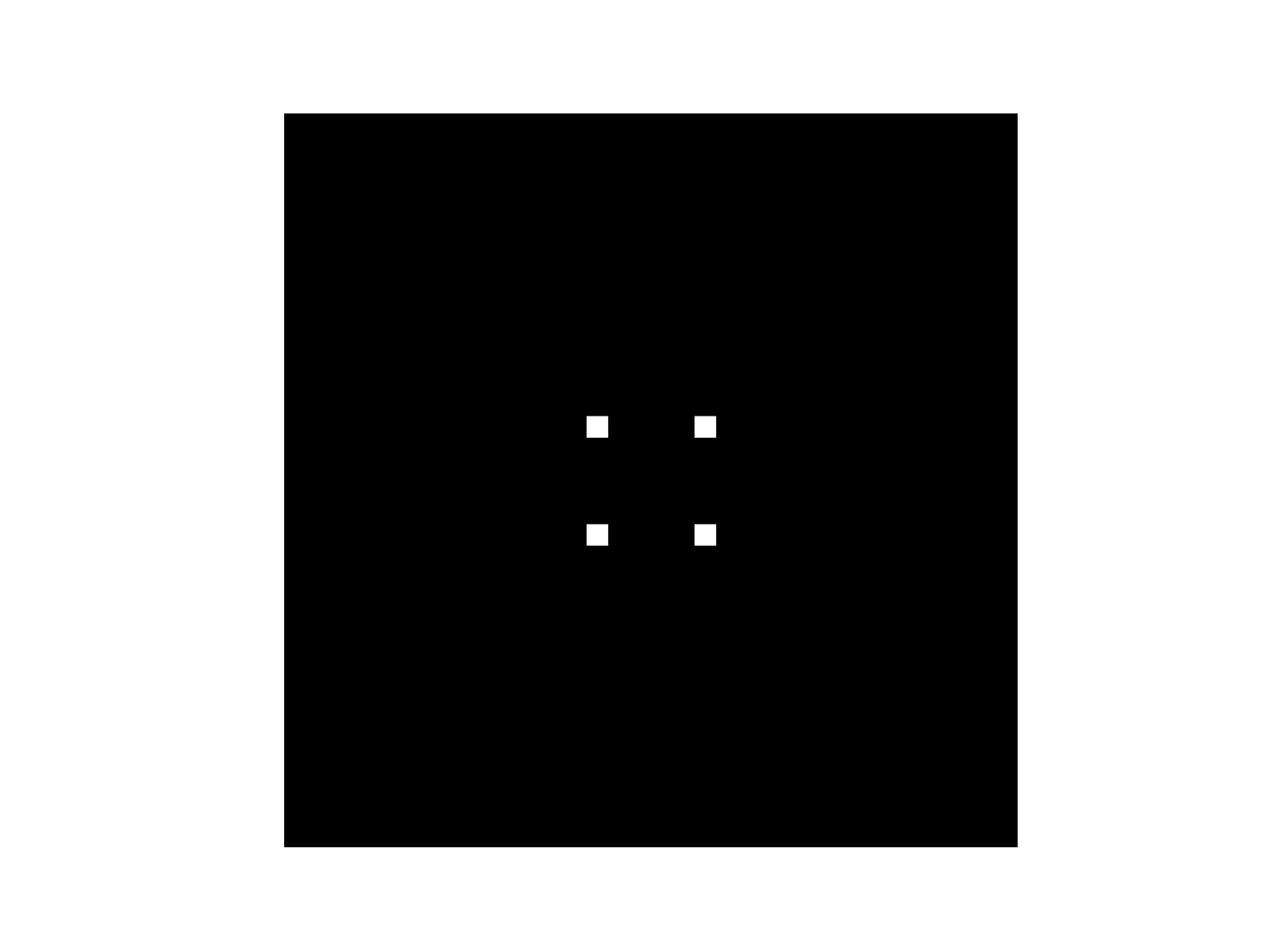}}
\qquad
\subfigure[]{
\label{fig:centerPokePhase}
\includegraphics[height=2.25in]{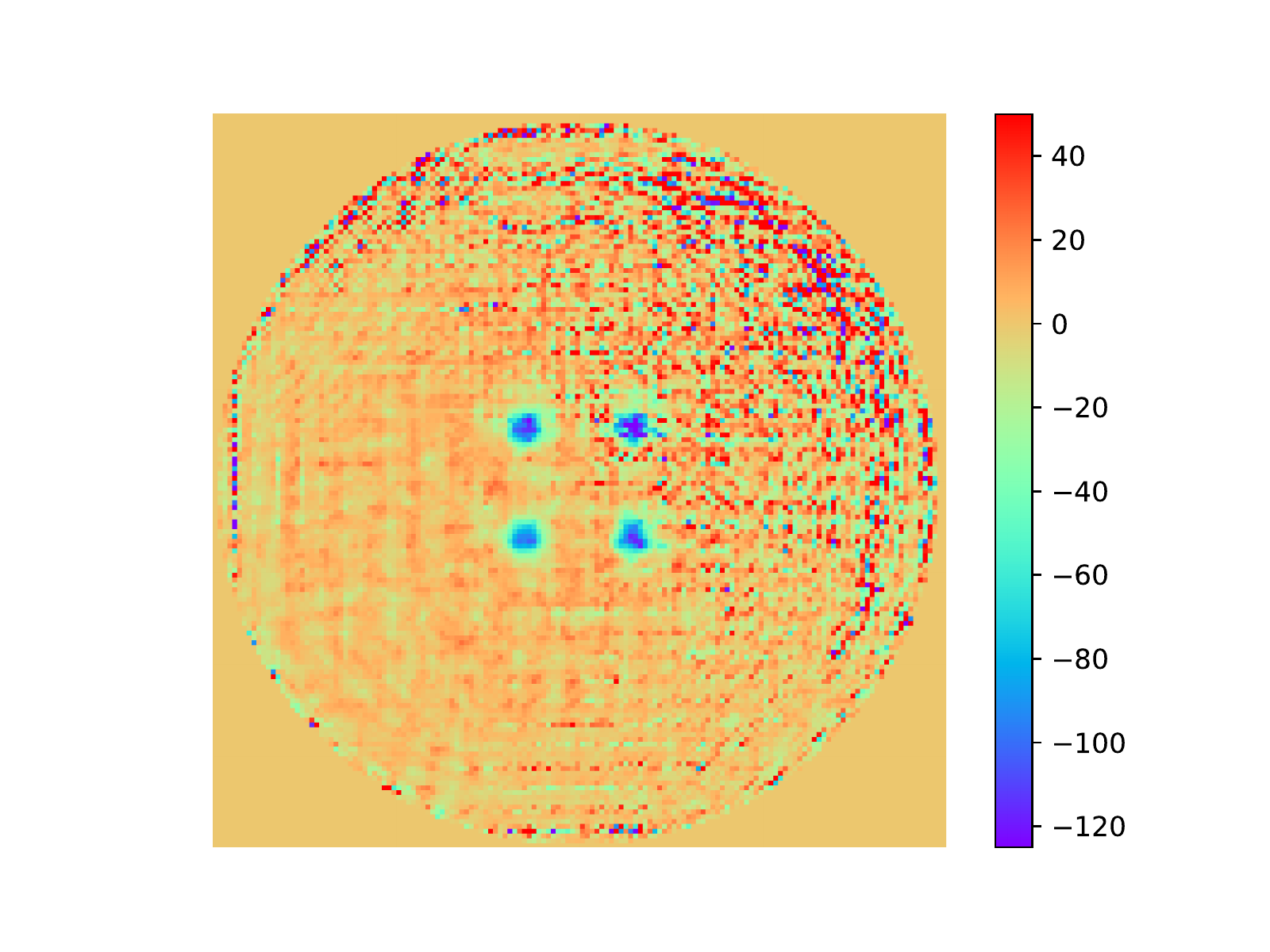}}
\caption{(a) DM command and (b) retrieved phase for the case where four known actuators are poked to determine the location of the center of the DM aperture with respect to the phase retrieval phase map.  Color scale is in nm.
\label{fig:centerPoke} }
\end{figure}

The second calibration step is to determine the location of the center of the DM in the array of data calculated by the phase retrieval algorithm.  To facilitate this, four actuators around the center of the DM are poked, and the associated phase is retrieved.  This is illustrated in \autoref{fig:centerPoke}.  The intersection of two diagonal lines connecting the actuators unambiguously locates the center of the DM.

\begin{figure}[h]
\centering
\includegraphics[width=0.75\textwidth]{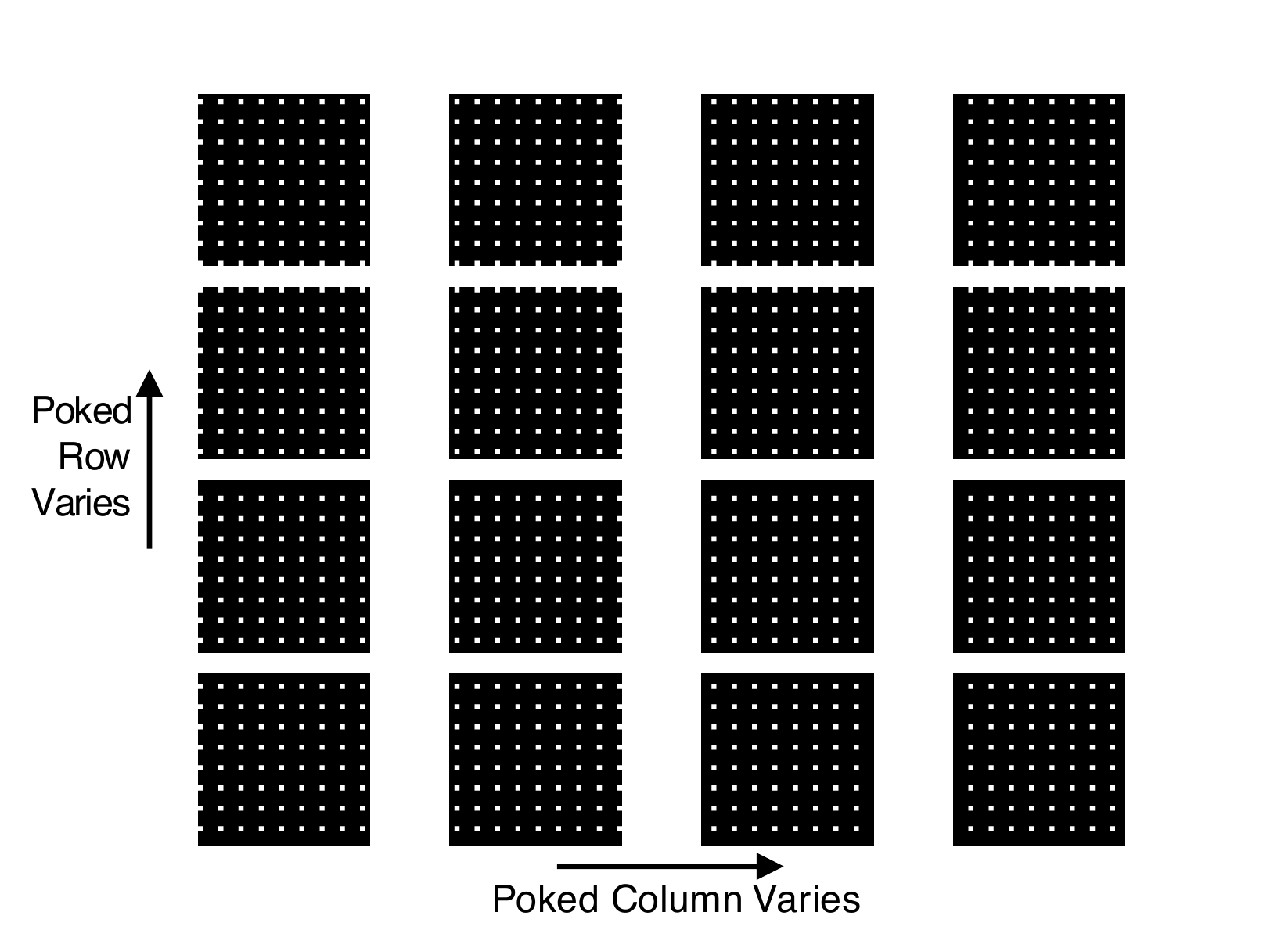}
\caption{Sixteen checkerboard DM commands which drive, in turn, all sixteen actuators in a four by four cell. Note the shift of which actuator row and column is active between each of the 16 commands. }
\label{fig:DM_Checkerboard}
\end{figure}

\begin{figure}[h]
\centering
\includegraphics[width=0.7\textwidth]{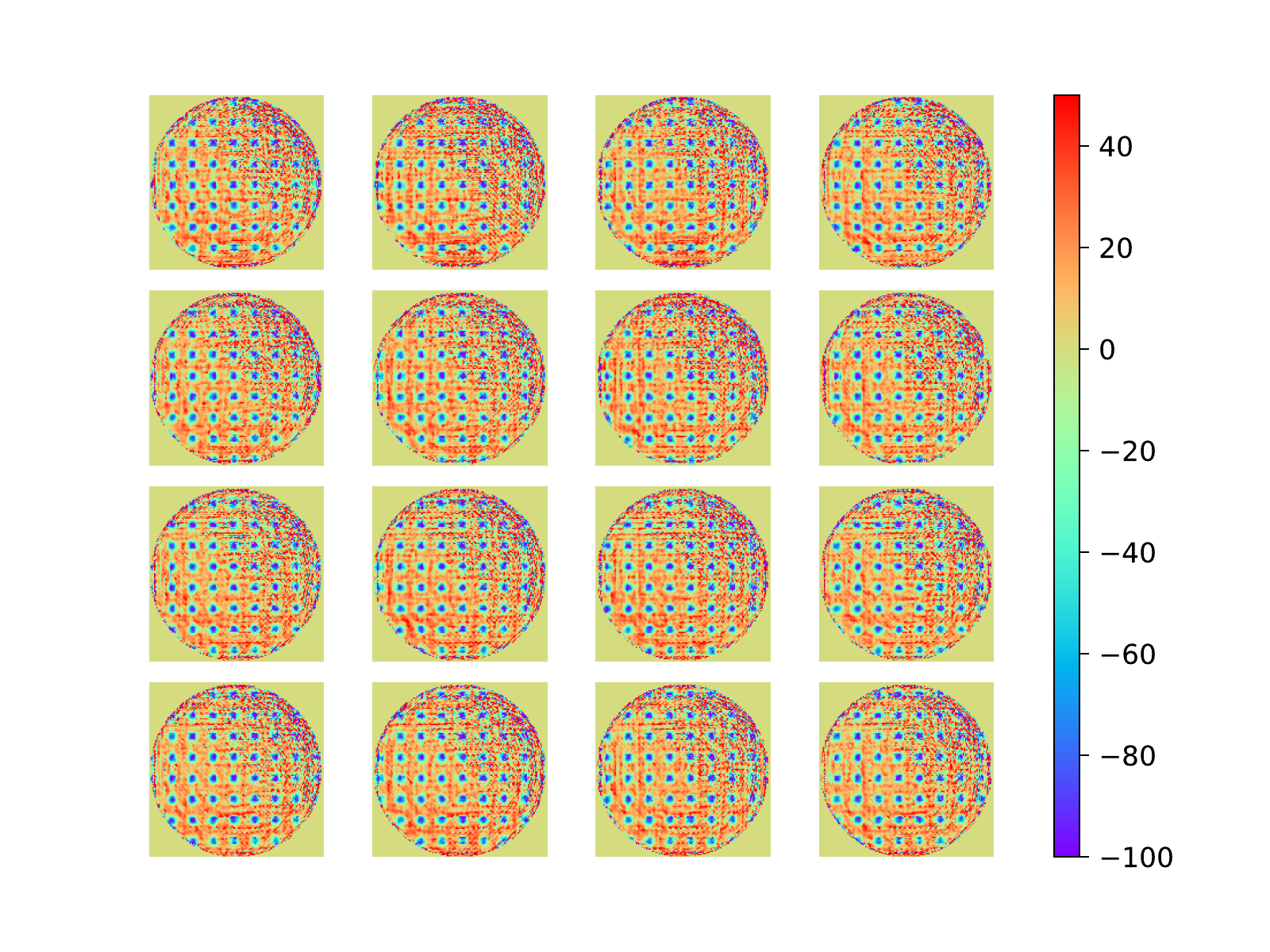}
\caption{Retrieved phase maps for the 16 checkerboard patterns applied to the DM.  Color scale is in nm.
\label{fig:Phase_Checkerboard} }
\end{figure}

Then, the mapping of the recovered wavefront to the individual DM actuators is determined.  This is done by putting a ``checkerboard" pattern on the DM in which every fourth actuator is poked, illustrated in \autoref{fig:DM_Checkerboard}. Sixteen data sets were recorded where each actuator in a $4 \times 4$ cell is poked.  From this, the location of each actuator can be unambiguously determined in the array of wavefront data by comparison with the corresponding phase retrieval result.  The phase retrieval results corresponding to \autoref{fig:DM_Checkerboard} are shown in \autoref{fig:Phase_Checkerboard}.  Mapping the pupil in this way calibrates for any pupil imaging distortion between the DM plane and the exit pupil of the system.

\subsection{Uncorrected System Wavefront Measurement}
\begin{figure}[H]
\centering
\includegraphics[height=2.8in]{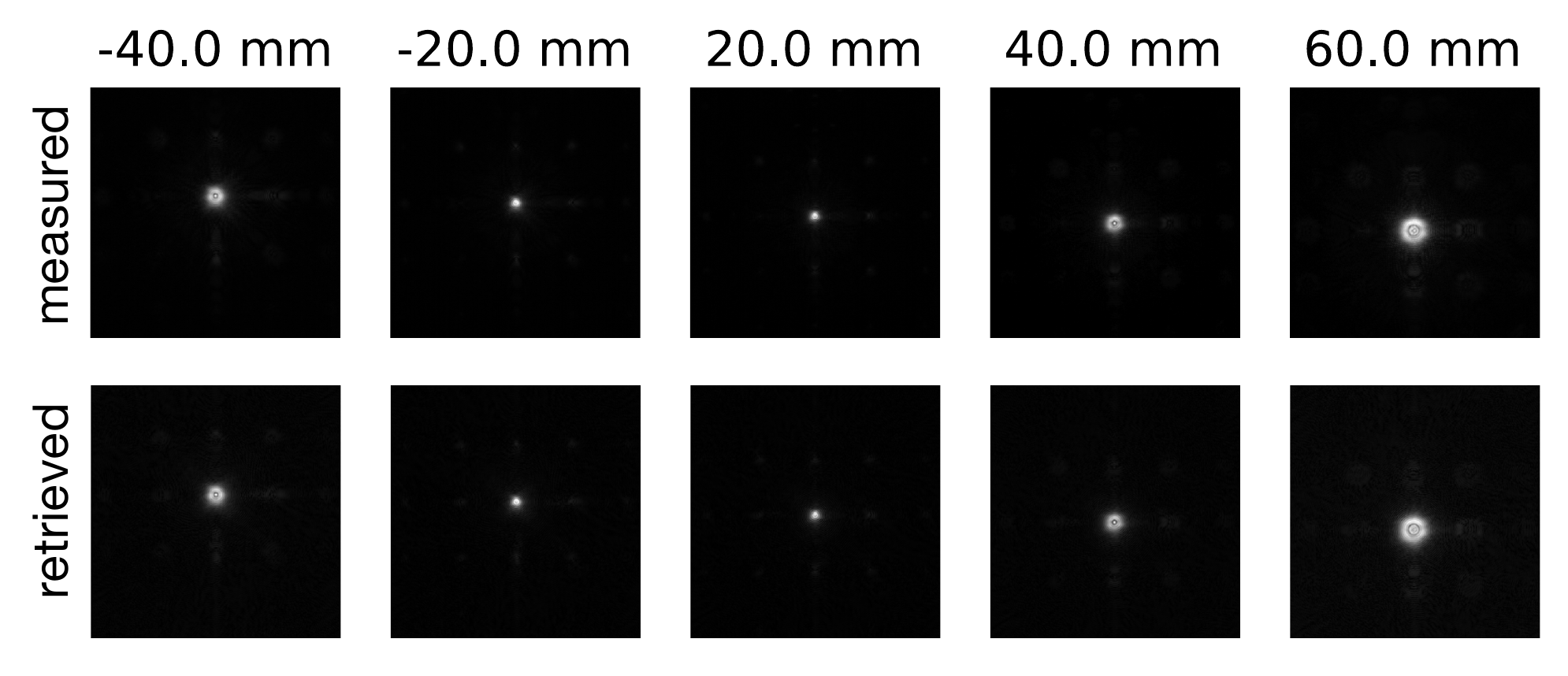}
\caption{Intensity patterns measured in the system before any wavefront correction was performed using phase retrieval.  The images in the top row are the minimally processed intensity measurements from the phase retrieval camera.  The images in the bottom row are the intensity distributions modeled by our phase retrieval algorithm using the recovered wavefront.  The distances labeled across the top of the figure indicate the axial position of measurement plane with respect to the nominal focus plane.
\label{fig:UncorrPSFs} }
\end{figure}

\begin{figure}[H]
\centering
\includegraphics[height=3.0in]{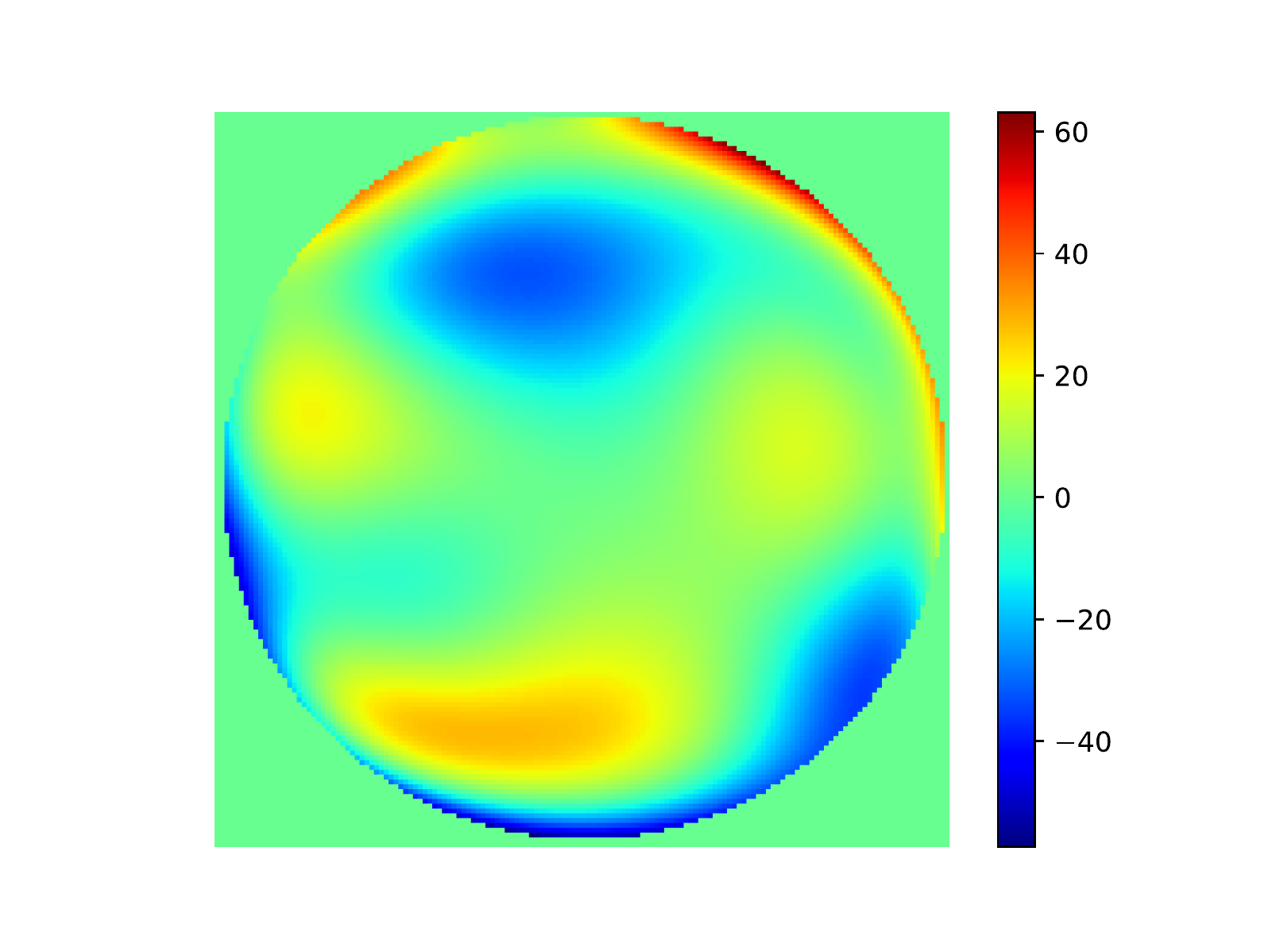}
\caption{Initial wavefront in system exit pupil before applying corrections.  The PV wavefront error is 120 nm and the RMS wavefront error is 16 nm.  The color scale is in units of nm.
\label{fig:UncorrWavefront} }
\end{figure}

In order to measure the system wavefront, we first applied our best-known flat maps to DM1 and DM2, determined before their installation using a Fizeau interferometer.  Thus the system was as well-corrected as possible without any additional correction using phase retrieval.  The recorded intensity data is shown in the top row of \autoref{fig:UncorrPSFs}.  Each of these images is the sum of 200 registered images.  This summing process results in an image with high dynamic range and excellent SNR.  The phase recovered from this data using the first 45 Zernike polynomials is shown in  \autoref{fig:UncorrWavefront} and the corresponding modeled intensity distributions are shown in the bottom row of \autoref{fig:UncorrPSFs}.  The Peak-to-Valley (PV) wavefront error is 120 nm and the Root-Mean-Squared (RMS) wavefront error is 16 nm.

\subsection{System Wavefront and Point-Spread Function with Corrections Applied}
\begin{figure}[H]
\centering
\includegraphics[height=2.8in]{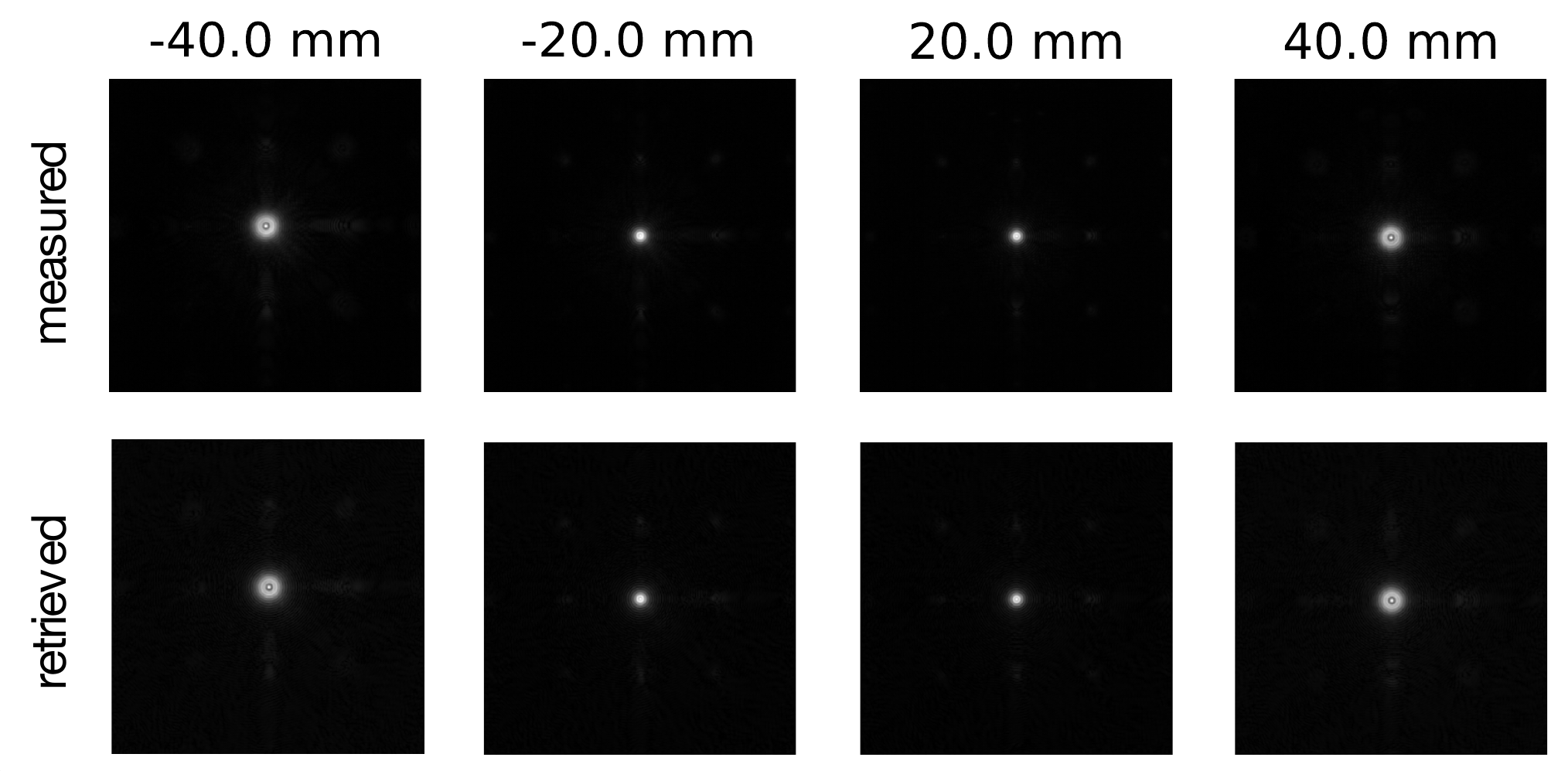}
\caption{Intensity patterns measured in the system after a correction based on the wavefront shown in \autoref{fig:UncorrWavefront} was applied.  The images in the top row are the minimally processed intensity measurements from the phase retrieval camera.  The images in the bottom row are the intensity distributions modeled by our phase retrieval algorithm using the recovered wavefront.  The distances labeled across the top of the figure indicate the axial position of measurement plane with respect to the nominal focus plane.
\label{fig:CorrPSFs22mm} }
\end{figure}

\begin{figure}[H]
\centering
\includegraphics[height=3.0in]{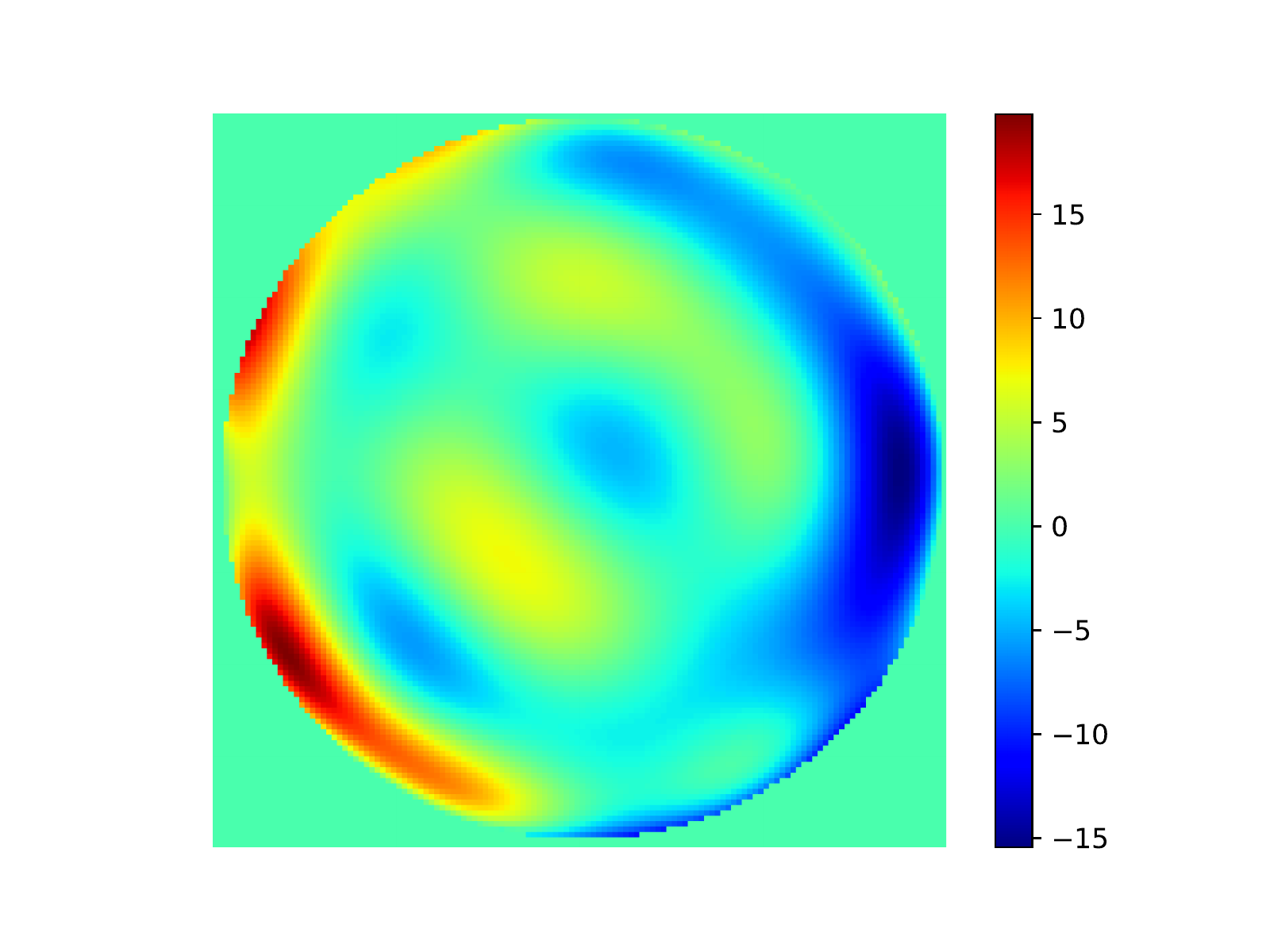}
\caption{Corrected system wavefront, using correction derived from data in \autoref{fig:UncorrWavefront}, in exit pupil of system in the case where an oversized, 22 mm diameter aperture was used.  The PV wavefront error is 35.2 nm and the RMS wavefront error is 5.5 nm.  The color scale is in units of nm.
\label{fig:CorrWavefront22mm} }
\end{figure}

The conjugate of the wavefront in \autoref{fig:UncorrWavefront} was applied to the DM using the calibration and mapping determined above.  The recorded PSF data is shown in the top row of \autoref{fig:CorrPSFs22mm}.  The phase recovered from this data using the first 45 Zernike polynomials is shown in \autoref{fig:CorrWavefront22mm}.  The PV wavefront error is 35.2 nm and the RMS wavefront error is 5.5 nm.  This is a significant improvement over the uncorrected case.  This correction is a strong indication of the precision and accuracy of our phase retrieval results and demonstrates the validity of our methodology for applying the correction to the DM.

\subsection{Final Wavefront Over 18 mm Aperture}
\begin{figure}[H]
\centering
\includegraphics[height=2.8in]{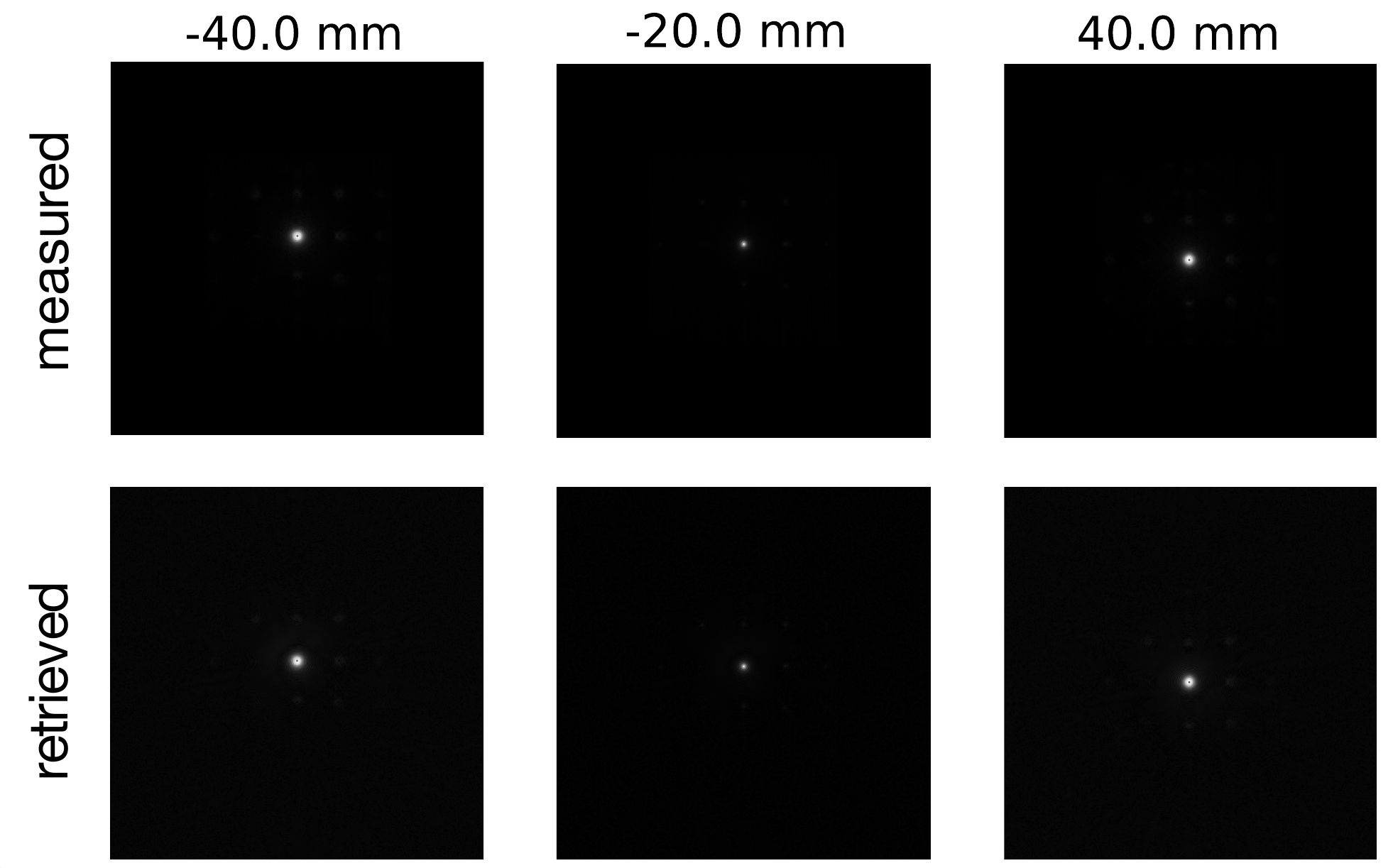}
\caption{Corrected PSFs, measured and retrieved over 18 mm aperture
\label{fig:CorrPSFs18mm} }
\end{figure}

\begin{figure}[H]
\centering
\includegraphics[height=3.0in]{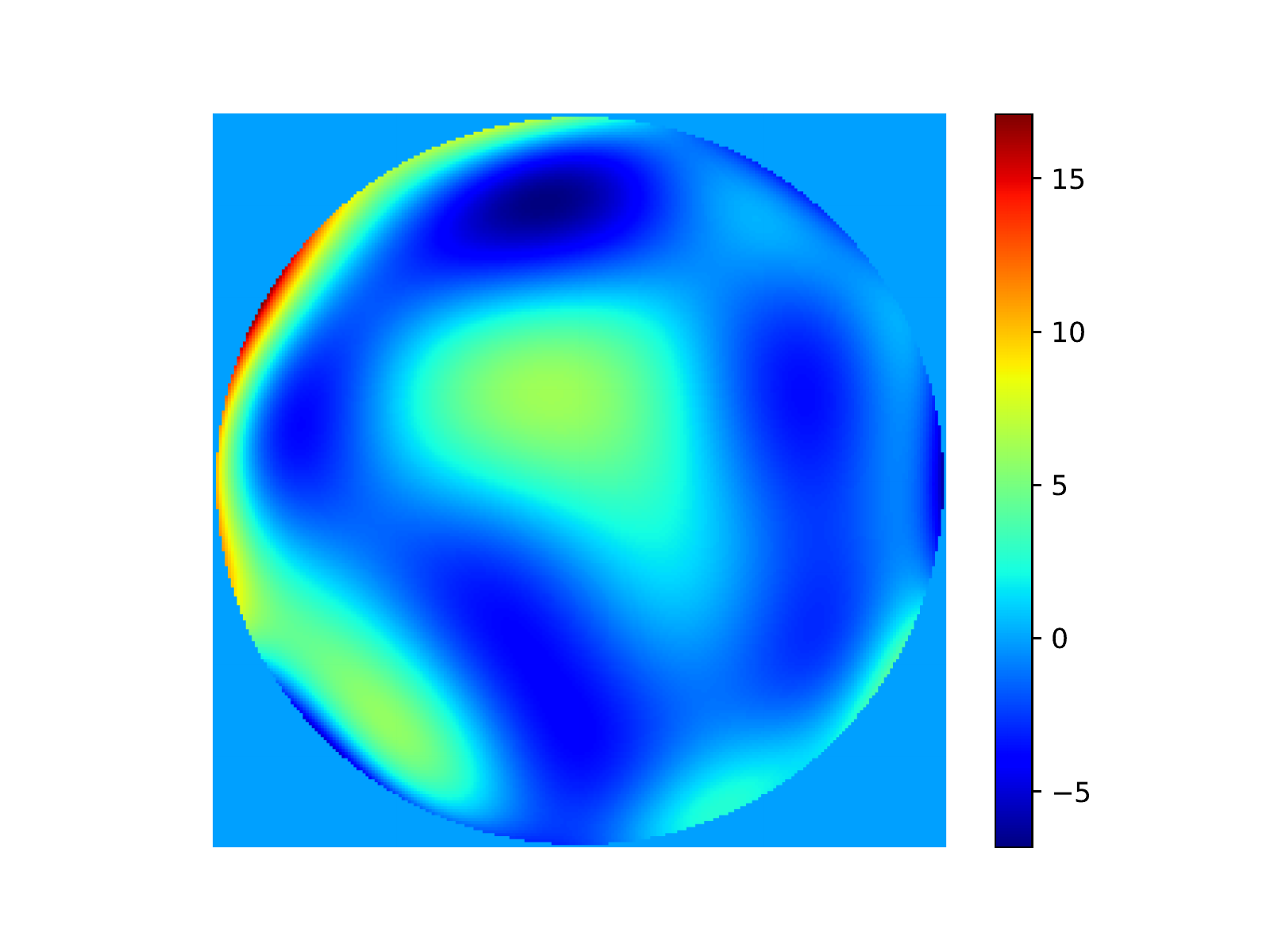}
\caption{Corrected wavefront over system's nominal 18 mm aperture.  The PV wavefront error is 23.9 nm and the RMS wavefront error is 3.0 nm.  The color scale is in units of nm.
\label{fig:CorrWavefront18mm} }
\end{figure}

\begin{figure}[H]
\centering
\includegraphics[width=4.0in]{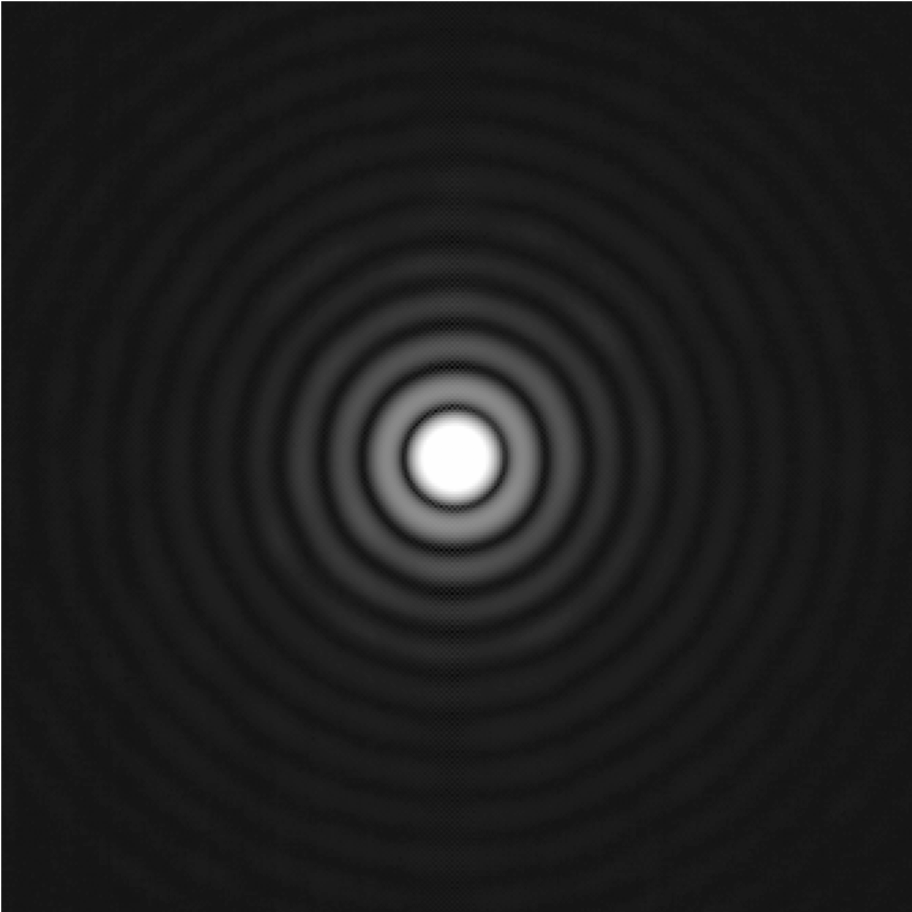}
\caption{In-focus PSF over 18 mm aperture.  The data here is the sum of 1000 registered camera frames, resulting in a very low SNR image with high dynamic range.  The data is plotted on a logarithmic scale with a compressed dynamic range so that very dim outer rings are visible.  The PSF is almost indistinguishable from a perfect theoretical result for a circular pupil with no aberrations and, qualitatively, is consistent with our measured wavefront error of 3 nm RMS.
\label{fig:CorrInfocusPSF18mm} }
\end{figure}

After the first wavefront correction, the oversized 22 mm aperture was removed and the nominal 18 mm pupil aperture was inserted.  A subsequent data set was taken.  Example intensity images are shown in the top row of \autoref{fig:CorrPSFs18mm}.  The retrieved wavefront is shown in \autoref{fig:CorrWavefront18mm}.  The PV wavefront error is 23.9 nm and the RMS wavefront error is 3.0 nm.  An image of the in-focus PSF of the system, resulting from the average of 1000 images is shown in \autoref{fig:CorrInfocusPSF18mm}.  It is difficult to distinguish this from a plot of a perfect theoretical result.

\section{Summary}

We have described the parametric approach to phase retrieval, including details of our forward model and description of the parameters and corresponding derivatives used as inputs to an optimization algorithm.  We then described its application to the measurement of the transmitted wavefront error of a coronagraphic testbed system and its subsequent correction using a deformable mirror.  We included details of the calibrations necessary to register the measured wavefront maps to the DM actuators, including confirming the orientation and polarity of the data, determining the center point of the DM and calibrating for distortions.  The measured wavefront error improved from 16 nm RMS before correction to 3.0 nm RMS after correction, which is summarized in \autoref{table:summaryTable}.  The fact that we are able to measure a low amplitude wavefront, derive and apply a correction from this measurement, and realize a significantly improved transmitted wavefront, demonstrates the precision and accuracy of our phase retrieval algorithm and correction methodology.

\begin{table}[H]
 \caption{Summary of wavefront errors before and after correction}
 \label{table:summaryTable}
\begin{center}
 \begin{tabular}
 {||c c||} 
 \hline
 \textbf{Case} & \textbf{RMS Wavefront Error (nm)} \\ [0.5ex] 
 \hline\hline
 Initial Aligned System (DM1 and DM2 flattened)& 16  \\ 
 \hline
 Correction Applied over 22 mm Aperture & 5.5  \\
 \hline
 Correction Applied over 18 mm Aperture & 3.0 \\
 \hline
\end{tabular}
\end{center}
\end{table}

\acknowledgments 
This work is supported by the National Aeronautics and Space Administration under Grants NNX12AG05G, NNX14AD33G issued through the Astrophysics Research and Analysis (APRA) program (PI: R. Soummer) the JWST Telescope Scientist Investigation, NASA Grant NNX07AR82G (PI: C. Matt Mountain) and the STScI Director’s Discretionary Research Funds. 

\bibliography{references}
\bibliographystyle{spiebib}

\end{document}